\documentclass[preprintnumbers, prd, showpacs, onecolumn, floatfix, preprintnumbers, letterpaper, amsmath, amssymb, superscriptaddress]{revtex4}

\usepackage{amssymb, amsmath, bm, dcolumn, epsf, graphicx, latexsym, mathbbol, slashed, simplewick}

\usepackage{color}
\newcommand{\tc}{\textcolor{black}}

\newcommand{\tb}{\textcolor{black}}

\newcommand{\be}{\begin{equation}}
\newcommand{\ee}{\end{equation}}
\newcommand{\bea}{\begin{eqnarray}}
\newcommand{\eea}{\end{eqnarray}}
\newcommand{\barr}{\begin{array}}
\newcommand{\earr}{\end{array}}

\begin{document}

\hfill USTC-ICTS-14-06
\title{Detecting relic gravitational waves in the CMB: The contamination caused by the cosmological birefringence}

\author{Wen Zhao}
\affiliation{Key Laboratory for Researches in Galaxies and Cosmology, Department of Astronomy, University of Science and Technology of China, Hefei, Anhui,
230026, China}

\author{Mingzhe Li}
\affiliation{Interdisciplinary Center for Theoretical Study, University of Science and Technology of China, Hefei, Anhui 230026, China}


\begin{abstract}
{The B-mode polarization of the cosmic microwave background (CMB) radiation is an excellent information channel for the detection of relic gravitational waves. However, the detection is contaminated by the B-mode polarization generated by some other effects. In this paper, we discuss the contaminations caused by the cosmological birefringence, which converts the CMB E-mode to the B-mode, and forms the effective noise for the detection of gravitational waves. We find that this contamination is significant, if the rotation angle is large. However, this kind of B-mode can be properly de-rotated, and the effective noises can be greatly reduced. We find that, comparing with the contaminations caused by cosmic weak lensing, the residual polarization generated by the cosmological birefringence is negligible for the detection of relic gravitational waves in the CMB. }
\end{abstract}

\maketitle

\section{Introduction}

The temperature and polarization anisotropies of the cosmic microwave background (CMB) radiation contain fruitful cosmological information, and play crucial roles in the determinations of various parameters in modern cosmology. In particular, the CMB anisotropies provide the unique observational way to detect the relic (primordial) gravitational waves, which was inevitably produced in the very early Universe \cite{grishchuk1979,starobinsky1980}. Relic gravitational waves left the observable imprints in all the CMB power spectra, including the TT, TE, EE and BB information channels. Limited by the cosmic variance, if the tensor-to-scalar ratio $r<0.05$ \cite{turner1994,zhao2009}, the TT, TE and EE channels become useless, and we can only detect it in the B-mode polarization \cite{zaldarriaga1997,kamionkowski1997}, which is one of the most important goals for the CMB efforts \cite{task,krauss2010}. {\tb {The current satellite observations, including those of the WMAP \cite{wmap9} and Planck missions \cite{planck}, are yet to detect a definite signal of relic gravitational waves. However, the recent observations of the ground-based experiments, such as the BICEP1, SPTPOL, POLARBEAR, ACTPOL telescopes, have given some interesting results for the B-mode polarizations \cite{bicep1,sptpol,polarbear,actpol}. In particular, the BICEP2 team released their recent data, and claimed the discovery of relic gravitational waves with the tensor-to-scalar ratio $r=0.20^{+0.07}_{-0.05}$, and $r=0$ disfavored at $7.0\sigma$ \cite{bicep2}. However, it has been shown that this observation is also consistent with the polarized radiation emitted by the poorly-understood interstellar dust \cite{debate1,debate2,liuhao}. In order to distinguish them, one should measure the B-mode polarization at the different frequency channels by BICEP2 or some other experiments (such as the Planck satellite, BICEP3, EBEX, QUBIC and so on). Nevertheless, these results encourage us to put the gravitational-wave detection through the CMB polarization as a highest priority for the next generations of the CMB experiments \cite{task,krauss2010}. In addition, we should mention that it is important to cross-check the CMB results by the other large-scale observations. For example, in the previous works \cite{c1,c2}, the authors found that the relic gravitational waves can distort the shapes of galaxies through the effects of tidal fields of the large-scale structure and the gravitational lensing, and form the B-modes in galaxy shape correlations, which can also be used to detect the signal of gravitational waves \cite{d1,c3}. In \cite{c3}, the authors discussed in detail the possibility of the gravitational-wave detection by using the cross-correlation between shear and CMB B-mode polarization, and found that it is quite possible to confirm or falsify the BICEP2 results, if the next-generation surveys beyond EUCLID, WFIRST, and LSST are considered.
}}

In the standard cosmological model, up to the first-order perturbations, {\tc {both the vector and tensor perturbations \cite{zaldarriaga1997,kamionkowski1997} have the possibilities to generate CMB B-mode. However the vector perturbations usually decay quickly in the expanding universe and can be ignored in CMB physics. Hence significant primordial CMB B-mode can only be generated by the tensor perturbations, i.e., the relic gravitational waves. In principle, such signals for the relic gravitational waves}} cannot be polluted by the primordial density perturbations, which is the reason why the method of detecting gravitational waves through CMB B-mode is clean. However, in the real Universe, the detection ability of the method is limited by various contaminations. In addition to the foreground radiations, various systematic errors, the E-B mixture caused by the incomplete sky survey, and the instrumental noises in the real observations, some other effects can also naturally generated the B-mode polarization. The well-studied one is the cosmic weak lensing, which mixed the E-mode and B-mode polarizations, and forms a nearly multipole-invariant BB power spectrum \cite{zaldarriaga1998,lewis2006}.

Another effect to produce the B-mode polarization is the so-called cosmological birefringence, which can be caused by the possible coupling between the electromagnetic field and the scalar field (which may or may not be identified as the dark energy) through the Chern-Simons term \cite{carroll,kamionkowski1998,other-cpt,li2009}. The cosmological birefringence generates a frequency-independent rotation of the linear polarization of the CMB photons when they propagate over the cosmological distances. So the B-mode polarization can be naturally converted from the E-mode, even if it is absent in the early universe. This phenomenon resulted from the Lorentz and $CPT$ violations in the electrodynamics provides an effective method to test fundamental symmetries of nature and attracted many interests. {\tc {It is similar to but different from the Faraday rotation caused by the cosmological magnetic field \cite{Faraday1,Faraday2,Faraday3}, where the rotation of the polarization depends on the frequency of the photon. In this paper the ``cosmological birefringence" only means the frequency-independent rotation, as in Refs. \cite{carroll,kamionkowski1998}. For this phenomenon,}} the rotation of the CMB photons is only quantified by the rotation angle $\alpha$. Numerous works have constrained it by using the current CMB data, and show some evidences of the nonzero result \cite{feng2006,other-constraints,xia2012}. The current tightest constraint comes from the data analysis in \cite{xia2012}, where the author found that ${\alpha}=-2.28\pm 1.02  ~{\rm deg}$ (1$\sigma$) when considering the seven-year WMAP, BOOMERanG 2003 and BICEP data. This follows that $|\alpha|<4.32 ~{\rm degree}$ in the $2\sigma$ confidence level. However, we should remember that this result is still in debate. For instance, in the same paper, the author also found that $-1.34 < \alpha < 0.82~{\rm degree}$ at $95\%$ confidence level, if adding the QUaD polarization data.

In this paper, we shall investigate the CMB B-mode polarization produced by the cosmological birefringence, and focus on its influence on the detection of relic gravitational waves. We find that this B-mode could be quite large, if the rotation angle is close to the current upper limit value. If considering this B-mode as a new effective noise, the gravitational-wave detection in the CMB is impossible when $r<0.0014$. So, it is important to remove this contamination for the future observations. In our discussion, we propose a method to de-rotate it by utilizing the statistical properties of the E-mode polarization, and those of the estimator of $\alpha$ parameter. We find that, if considering the de-rotating, the residual B-mode polarization becomes very small. Comparing with contamination of the cosmic weak lensing, the residuals become negligible for the detection of gravitational waves.

The outline of this paper is as follows.
In Sec. \ref{sec2}, we briefly introduce the cosmological birefringence and the rotated CMB linear polarizations.
In Sec. \ref{sec3}, we discuss the method to de-rotate the CMB B-mode polarization, and the influence on the gravitation-wave detection.
In Sec. \ref{sec4}, we summarize the main results of this paper.

\section{CMB polarizations and the cosmological birefringence \label{sec2}}

The CMB linear polarization can be described by the Stokes parameters $Q$ and $U$. In general, these two fields on the sky can be expanded as follows,
 \begin{equation}
 (Q\pm iU)(\hat{\bf n}) = \sum_{\ell m} (E_{\ell m}\pm iB_{\ell m}) ~_{\pm 2}Y_{\ell m}(\hat{\bf n}),
 \end{equation}
where $_{\pm 2}Y_{\ell m}(\hat{\bf n})$ are the spin-weighted spherical harmonics and we have already considered the E/B decomposition. The power spectra are defined by
 \begin{equation}
 C_{\ell}^{EE}=\frac{1}{2\ell+1}\sum_{m}\langle E_{\ell m} E_{\ell m}^*\rangle, ~~C_{\ell}^{BB}=\frac{1}{2\ell+1}\sum_{m}\langle B_{\ell m} B_{\ell m}^*\rangle,
 \end{equation}
where the brackets denote the average over all realizations. Under the assumption of Gaussian and the statistically isotropic fields, the statistical properties of the CMB maps are specified fully by these polarization spectra EE and BB, auto-correlation of CMB temperature anisotropy TT, and their cross-correlations TE, TB and EB. Note that, in the cases without the cosmological birefringence, $C_{\ell}^{TB}=C_{\ell}^{EB}=0$ due to the parity symmetry of the universe. {\tc {Besides the cosmological birefringence, non-zero TB and EB correlations can also be produced by the Faraday rotation when the CMB photons pass through a cosmological magnetic field with a non-zero helicity \cite{Faraday2}, which sets a special direction in the universe and spontaneously breaks the spatial isotropy as well as the parity symmetry along this direction. This case is beyond the scope of this paper.}}

Now, let us consider the Chern-Simons coupling between the scalar field $\varphi$ and the CMB photons
 \begin{equation}\label{chern}
 \mathcal{L}_{int}=\frac{\beta\varphi}{2M}F^{\mu\nu}\tilde{F}_{\mu\nu},
 \end{equation}
where $F_{\mu\nu}$ is the electromagnetic field-strength tensor and $\tilde{F}^{\mu\nu}=1/2 \epsilon^{\mu\nu\rho\sigma}F_{\rho\sigma}$ is its dual, $\beta$ is the dimensionless coupling constant and $M$ is the new energy scale of the theory. In this paper, we will not consider the fluctuations of the scalar field \cite{li2008,cpt-perturbation,kamionkowski2009}, which has be discussed in the separate paper \cite{zhao2014}. We assume that the scalar field is spatially homogeneous but changing with time. When the CMB photons propagate from the last scattering surface (LSS) to us, their polarization planes are rotated by an angle $\alpha$ through the Chern-Simons term due to the Lorentz and $CPT$ violations \cite{li2008},
\be
  (Q\pm iU)^{\rm rd}(\hat{\bf n})=e^{\pm 2i\alpha} (Q\pm iU)(\hat{\bf n})~,
  \ee
  where the superscript ${\rm rd}$ denotes the rotated variables and the variables without it are those if the cosmological birefringence is absent.
  The rotation angle $\alpha$ is given by $\alpha=\beta \Delta\varphi/M$ \cite{carroll1990,li2008}, and $\Delta \varphi=\varphi_0-\varphi_{LSS}$ is the change of $\varphi$ from  the LSS to the present time.
Correspondingly the rotated E-mode and B-mode coefficients become
 \begin{equation}\label{eb-rd1}
 E_{\ell m}^{\rm rd}=\cos(2\alpha)E_{\ell m}-\sin(2\alpha)B_{\ell m},~~~B_{\ell m}^{\rm rd}=\sin(2\alpha)E_{\ell m}+\cos(2\alpha)B_{\ell m}.
 \end{equation}
So, except the TT spectrum all other CMB power spectra change as
\bea\label{rotationformulas}
  C^{TE,{\rm rd}}_{\ell}&=&C^{TE}_{\ell} \cos{(2\alpha)}~, \nonumber\\
 C^{TB,{\rm rd}}_{\ell}&=&C^{TE}_{\ell} \sin{(2\alpha)}~, \nonumber\\
 C^{EE,{\rm rd}}_{\ell}&=&C^{EE}_{\ell} \cos^2{(2\alpha)} +C^{BB}_{\ell} \sin^2{(2\alpha)}~, \nonumber\\
   C^{BB,{\rm rd}}_{\ell}&=&C^{EE}_{\ell} \sin^2{(2\alpha)} +C^{BB}_{\ell} \cos^2{(2\alpha)}~, \nonumber\\
  C^{EB,{\rm rd}}_{\ell}&=&\frac{1}{2}\sin{(4\alpha)} (C^{EE}_{\ell}-C^{BB}_{\ell}) ~.
  \eea
The full set of these formulae were first written down in Ref. \cite{feng2006} and used for detecting or constraining the cosmological birefringence in the data analysis.
In these formulae we have not included possible TB and EB correlations produced at early universe,  e.g., the asymmetric tensor perturbation generated during inflation through gravitational Chern-Simons term
would result non vanished TB and EB cross correlations on LSS \cite{kamionkowski1998}.
Relevant formulae for this more general case can be found in Ref. \cite{li2009}.
{\tc {It deserves pointing out that the model (\ref{chern}) is not the unique one to have the rotation effect described by Eqs.(\ref{rotationformulas}). Some other models, such as the model proposed in \cite{Myers:2003fd}, have the same effect. More general Lorentz violating models producing birefringence through different dimensional operators have been discussed in \cite{Kostelecky:2009zp}. Among these models, the Chern-Simons model (\ref{chern}) considered in this paper is the simplest and most studied one in the literature. }}

In this paper, we reasonably assume a small rotation angle, i.e., $|\alpha|\ll1$.
So the relations in Eq.(\ref{eb-rd1}) reduce to
 \begin{equation}\label{eb-rd2}
 E_{\ell m}^{\rm rd}=(1-2\alpha^2)E_{\ell m}-2\alpha B_{\ell m},~~~B_{\ell m}^{\rm rd}=2\alpha E_{\ell m}+(1-2\alpha^2)B_{\ell m},
 \end{equation}
and the rotated CMB power spectra become
 \begin{eqnarray}\label{spectra-rd1}
 C_{\ell}^{TE,{\rm rd}}&=&(1-2\alpha^2)C_{\ell}^{TE} ~, \nonumber\\
 C_{\ell}^{TB,{\rm rd}}&=&2\alpha C_{\ell}^{TE}~,\nonumber\\
 C_{\ell}^{EE,{\rm rd}}&=&(1-4\alpha^2)C_{\ell}^{EE}+4\alpha^2 C_{\ell}^{BB}~, \nonumber \\
 C_{\ell}^{BB,{\rm rd}}&=&(1-4\alpha^2)C_{\ell}^{BB}+4\alpha^2 C_{\ell}^{EE}~, \nonumber \\
 C_{\ell}^{EB,{\rm rd}}&=&2\alpha (C_{\ell}^{EE}-C_{\ell}^{BB})~.
  \end{eqnarray}

\begin{figure}
\begin{center}
\includegraphics[width=10cm]{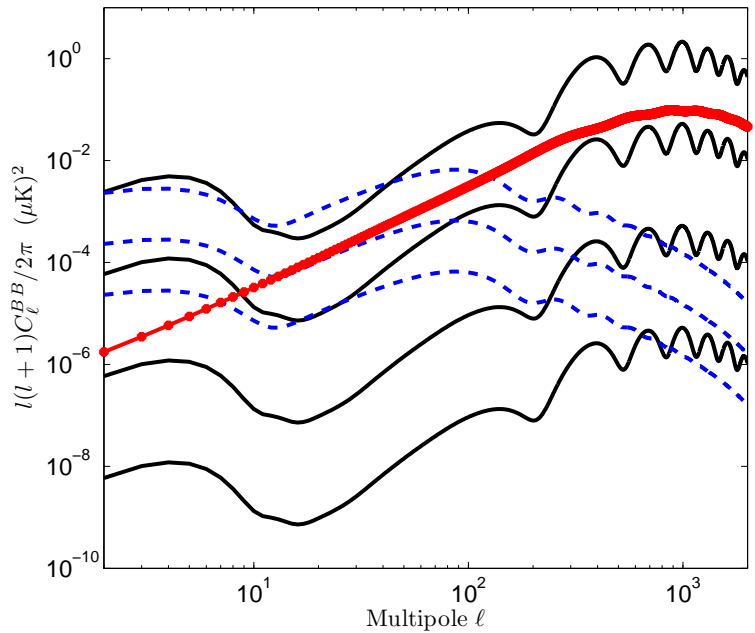}
\caption{\label{figb1} The black solid lines denote the B-mode power spectra generated by the cosmological birefringence effect. From the upper one to the lower one, we have considered the cases with the rotation angle $|\alpha|=4.32^{\circ}$, $1^{\circ}$, $0.1^{\circ}$, $0.01^{\circ}$, respectively. The blue dashed lines are the B-mode spectra generated by the primordial gravitational waves with the tensor-to-scalar ratio $r=0.1$ (upper), $0.01$ (middle), $0.001$ (lower). The red curve shows the B-mode caused by the cosmic weak lensing.}
\end{center}
\end{figure}

Now, let us focus on the B-mode power spectrum. Using Eq. (\ref{spectra-rd1}), we plot the BB spectrum in Fig. \ref{figb1}, where we have set the unrotated BB spectrum to zero. The amplitude of $C_{\ell}^{BB}$ is proportional to $\alpha^2$. If $|\alpha|=4.32^{\circ}$, the current 2$\sigma$ upper limit value, we find the rotated BB spectrum is quite large. It is interesting to compare it with the BB power spectrum generated by the relic gravitational waves. The recent Planck data give the constraint on the amplitude of gravitational waves $r<0.11$ \cite{planck}. From Fig.\ref{figb1}, we find the rotated $C_{\ell}^{BB}$ is larger than that of gravitational waves nearly in all the multipole range, even if the upper limit of the gravitational waves is considered.

\begin{figure}
\begin{center}
\includegraphics[width=10cm]{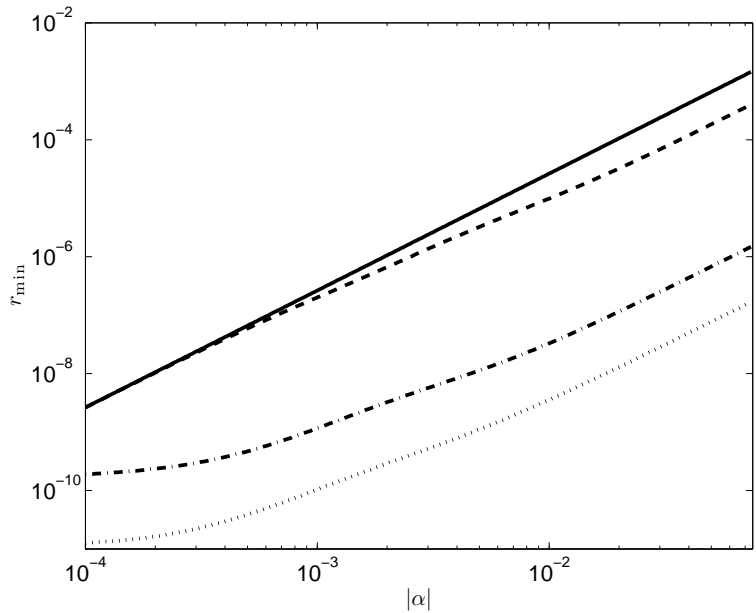}
\caption{\label{figb4} The detection limit of tensor-to-scalar ratio for the cases with different rotation angle $|\alpha|$. In all the cases, we have only considered the residual B-mode power spectrum generated the cosmological birefringence as the contaminations. The solid line denotes the result when we do not consider the de-rotating. The dashed line shows the results for the case, in which the de-rotating is proceeded by considering the Planck noises, the dash-dotted line is for the CMBPol noise case, and the dotted line is for the case of the reference experiment. }
\end{center}
\end{figure}

For the detection of relic gravitational waves in the CMB, the rotated B-mode power spectrum is an effective noise, which can limit the detection ability of the method. To quantify it, we define the signal-to-noise ratio for the detection of gravitational waves as follows \cite{zhao2009,zhao20092}
 \begin{equation}\label{snr}
 S/N=\sqrt{\sum_{\ell}\left(\frac{C_{\ell}^{BB}({g.w.})}{\Delta\hat{C}_{\ell}^{BB}(g.w.)}\right)^2},
 \end{equation}
where $C_{\ell}^{BB}(g.w.)$ is the BB spectrum generated by gravitational waves, and $\hat{C}_{\ell}^{BB}(g.w.)$ is the estimator. $\Delta\hat{C}_{\ell}^{BB}(g.w.)$ is the statistical uncertainty of the estimator, which can be approximated by
 \begin{equation}
 \Delta\hat{C}_{\ell}^{BB}(g.w.)=\sqrt{\frac{2}{(2\ell+1)f_{\rm sky}}}(C_{\ell}^{BB}(g.w.)+N_{\ell}^{BB}(g.w.)),
 \end{equation}
where $f_{\rm sky}$ is the sky-cut factor, and $N_{\ell}^{BB}(g.w.)$ includes all the effective noises in the detection. Here, we shall focus on the contamination cased by the rotated B-mode. So, we set $f_{\rm sky}=1$, i.e., a full-sky observation, and assume $N_{\ell}^{BB}(g.w.)=C_{\ell}^{BB,{\rm rd}}$, i.e., the contaminations come only from the rotated B-mode polarization. Given the values of the tensor-to-scalar ratio $r$ and the rotation angle $\alpha$, we can calculate the $S/N$ by using Eq.(\ref{snr}). For a fixed rotation angle, a larger $r$ follows a larger $S/N$. We define the quantity $r_{\rm min}$, which is the minimal $r$ value corresponds to a signal-to-noise ratio $S/N\ge 2$. So, the value of $r_{\rm min}$ stands for the detection limit of the method. In Fig.\ref{figb4}, we plot the $r_{\rm min}$ as a function of $\alpha$ (black line). For the case with $|\alpha|=4.32^{\circ}$, we have $r_{\rm min}=0.0014$, which means that if $r<0.0014$, the detection of gravitational waves become impossible due to the contamination caused by the cosmological birefringence. When $|\alpha|=1^{\circ}$, the detection limit becomes $r_{\rm min}=8\times 10^{-5}$, which is still a quite high limit.

\section{De-rotating and the residual B-mode polarization \label{sec3}}

To simplify the problem, in this section, we further assume the absence of the CMB B-mode polarization in the LSS. From the second formula in Eq.(\ref{eb-rd2}), we know that the rotated B-mode coefficients $B_{\ell m}^{\rm rd}$ depend only on the rotation angle $\alpha$, and the unrotated E-mode coefficients $E_{\ell m}$. So, if the values of $\alpha$ and $E_{\ell m}$ are known, one can completely reconstruct the B-mode polarization, and subtract it from the observable data. In the real observations, although none of them can be completely known from the observations, their statistical properties are well known. So, in principle, we can partly reconstruct the $B_{\ell m}^{\rm rd}$, and the residuals are expected to be much smaller. For the E-mode coefficients, we can easily construct their estimators $\hat{E}_{\ell m}$, which satisfy the Gaussian distribution with the zero expected value. Their variances are given by $\langle \hat{E}_{\ell m} \hat{E}^*_{\ell m}\rangle = C_{\ell}^{EE}+N_{\ell}^{EE}W_{\ell}^{-2}$, where $C_{\ell}^{EE}$ is the EE power spectrum, $N_{\ell}^{EE}$ is the instrumental noise power spectrum, and $W_{\ell}$ is the beam window function of the CMB detector.

For the rotation angle, we can also build the estimator $\hat{\alpha}$. The distribution function can be approximated as a Gaussian function. Since we anticipate the estimator is unbiased, we have the expected value $\langle \hat{\alpha}\rangle =\alpha$. The variance of the estimator can be determined from the real data analysis, which can be exactly obtained by using the Markov-chain Monte-Carlo likelihood analysis for the real data or mock data \cite{lewis2002}. In this paper, we approximate it by using the Fisher information matrix technique \cite{fisher}, which has been proved to be an excellent approximation for the determination of the observational uncertainties of the parameters, and been widely used in various parameter evaluations in cosmology. For the CMB case, the Fisher matrix is
 \begin{equation}
 F_{ij}=\sum_{\ell} \sum_{XX'} \sum_{YY'} \frac{\partial C_{\ell}^{XX'}}{\partial p_i} \Sigma^{-1}(\hat{C}_{\ell}^{XX'},\hat{C}_{\ell}^{YY'})
 \frac{\partial C_{\ell}^{YY'}}{\partial p_j},
 \end{equation}
where $p_i$ are the parameters to be determined, $XX'$ and $YY'$ can be TT, EE, BB, TE, TB, EB, depending on which information channel will be considered in the data analysis. The covariance matrix of the estimators is given by
 \begin{equation}
 \Sigma(\hat{C}_{\ell}^{XX'},\hat{C}_{\ell}^{YY'}) = \frac{\mathcal{C}_{\ell}^{XY}\mathcal{C}_{\ell}^{X'Y'}+
 \mathcal{C}_{\ell}^{XY'}\mathcal{C}_{\ell}^{X'Y}}{(2\ell+1)f_{\rm sky}},
 \end{equation}
where $\mathcal{C}_{\ell}^{XY}\equiv C_{\ell}^{XY}+N_{\ell}^{XY}W_{\ell}^{-2}$. Once the Fisher matrix is calculated, the variances of the parameter estimators can be evaluated by $(\Delta \hat{p}_i)^2\equiv\langle(\hat{p}_i-\langle \hat{p}_i\rangle)^2\rangle={F^{-1}}_{ii}$. Similar to the previous work \cite{kamionkowski2009}, in this paper, we only consider the TB and EB information channels, which dominate the contribution for the detection of cosmological birefringence in the CMB. In addition, only the rotation angle $\alpha$ is consider to be the parameter, which will be determined in the analysis. For the other cosmological parameters, we assume they have been well determined by the CMB channels TT, EE, BB and TE.

\begin{figure}
\begin{center}
\includegraphics[width=10cm]{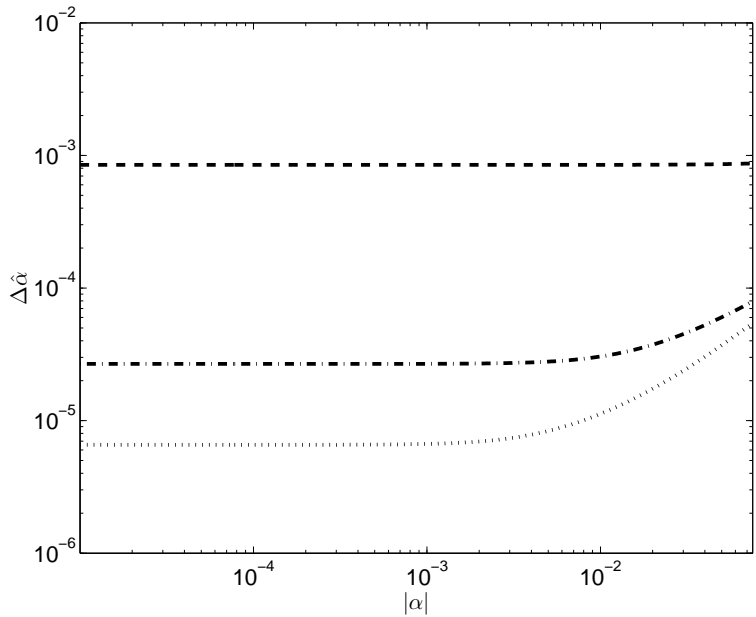}
\caption{\label{figb2} The uncertainties of the rotation angle $\Delta \hat{\alpha}$ as a function of $|\alpha|$. The black dashed line shows the results if the Planck noises are considered, the black dash-dotted line is for the CMBPol noise case, and the black dotted line is for the case of the reference experiment.}
\end{center}
\end{figure}

In the first case, we consider the noise level of the Planck satellite \cite{planck-bluebook}. The best frequency channel is at 143GHZ, in which the noise power spectra are \cite{ma2010} $N_{\ell}^{EE}=N_{\ell}^{BB}=2N_{\ell}^{TT}=2.79\times10^{-4}\mu{\rm K}^{2}$ for the 28-month survey. The beam full width at half maximum (FWHM) is $\theta_{\rm FWHM}=7.1'$, and the effective sky-cut factor is expected to be $f_{\rm sky}=0.65$. We plot the value of $\Delta\hat{\alpha}$ in Fig.\ref{figb2} (dashed line), which is $\Delta\hat{\alpha}\simeq 0.85\times10^{-3}$, nearly independent of the input rotation angle $\alpha$. As another example, let us consider the potential CMBPol mission, which is the fourth generation of the CMB experiments \cite{cmbpol}. For the best frequency channel at 150GHz. The noise power spectra are expected to be \cite{cmbpol,ma2010,zhao2011} $N_{\ell}^{EE}=N_{\ell}^{BB}=2N_{\ell}^{TT}=0.83\times10^{-6}\mu{\rm K}^{2}$, more than two orders smaller than those of Planck satellite. The beam FWHM is $\theta_{\rm FWHM}=5'$, and the effective sky-cut factor is $f_{\rm sky}=0.8$. For this case, we can calculate the values of $\Delta\hat{\alpha}$, which are also presented in Fig.\ref{figb2} with dash-dotted line. We find that, when the input angle $|\alpha|$ is smaller than 0.01, $\Delta\hat{\alpha}=2.7\times 10^{-5}$, nearly independent of the $\alpha$ value. However, if $|\alpha|>0.01$, the larger input $\alpha$ follows a larger $\Delta\hat{\alpha}$. When $|\alpha|=4.32^{\circ}$, the current upper limit value, we have $\Delta\hat{\alpha}=0.8\times10^{-4}$. In addition, similar to the previous work \cite{hu2002,knox2002}, we shall also consider a `reference' experiment as a far-future CMB observation. For this experiment, we assume the detector noise is $\Delta_p=\sqrt{2}\Delta_T=1\mu{\rm K}$-${\rm arcmin}$, which corresponds to the noise power spectra $N_{\ell}^{EE}=N_{\ell}^{BB}=2N_{\ell}^{TT}=0.85\times10^{-7}\mu{\rm K}^{2}$. The beam FWHM is assumed to be $1'$, and the sky-cut factor is assumed to be $1$. For this ideal case, the value of $\Delta\hat{\alpha}$ could be low as $0.7\times10^{-5}$, which can be found from the dotted line in Fig.\ref{figb2}.

Similar to the de-lensing method for the CMB polarization proposed in \cite{knox2002,kamionkowski2002}, we can also de-rotate the B-mode polarization caused by the cosmological birefringence. Given the reconstructed $\hat{\alpha}$ described above, and noisy observation of the E-mode $\hat{E}_{\ell m}$, we can define the estimator $\hat{B}_{\ell m}$ in a most general form,
 \begin{equation}\label{hatB}
 \hat{B}_{\ell m} = f(\ell,m)\hat{\alpha}\hat{E}_{\ell m},
 \end{equation}
where $f$ is a function of $\ell$ and $m$. Thus, the residual B-mode power spectrum is given by
 \begin{equation}
 C_{\ell}^{BB}({\rm residual})= \langle (B_{\ell m}^{\rm rd}-\hat{B}_{\ell m})({B_{\ell m}^{{\rm rd}*}}-\hat{B}^*_{\ell m}) \rangle.
 \end{equation}
We minimize the residual power spectrum, which determines the function $f$ in Eq.(\ref{hatB}) as follows
 \begin{equation}
 f(\ell,m)=2{\Theta}_{\ell},
 \end{equation}
where ${\Theta}_{\ell}\equiv\frac{\alpha^2}{\alpha^2+(\Delta\hat{\alpha})^2}\frac{C_{\ell}^{EE}}{C_{\ell}^{EE}+N_{\ell}^{EE}W_{\ell}^{-2}}$. The corresponding residual BB power spectrum is
 \begin{equation}
 C_{\ell}^{BB}({\rm residual})=4\alpha^2 C_{\ell}^{EE}(1-\Theta_{\ell}).
 \end{equation}

\begin{figure}
\begin{center}
\includegraphics[width=10cm]{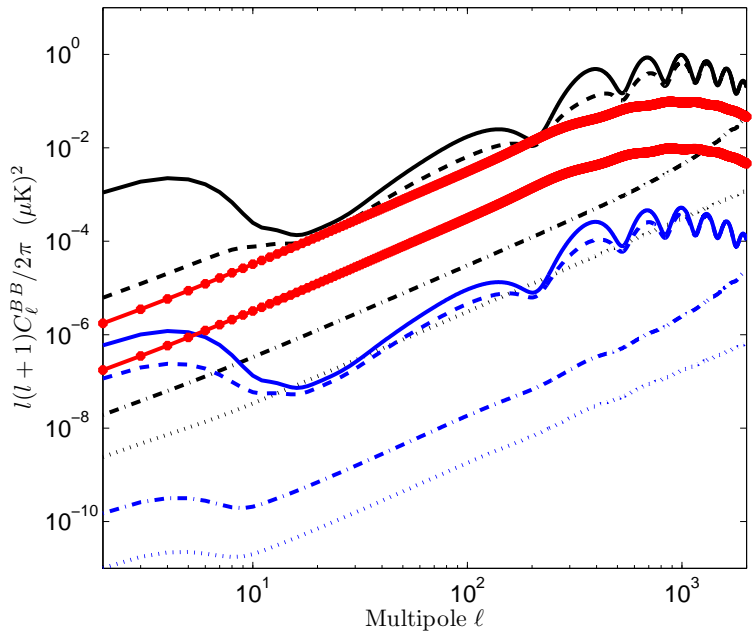}
\caption{\label{figb3} The black lines show the residual B-mode power spectra caused by the cosmological birefringence with the rotation angle $|\alpha|=4.32^{\circ}$. The solid line shows the case without de-rotating, the dashed line shows that the de-rotating is proceeded when considering the Planck noises, the dash-dotted line is the case when considering the CMBPol noises, and the dotted line is that for the noises of the reference experiment. The blue lines are exactly same with the black lines, but for the model with the rotation angle $|\alpha|=0.1^{\circ}$.
The red curve shows the original B-mode (upper line) and the de-lensed B-mode (lower line) power spectra caused by the cosmic weak lensing.}
\end{center}
\end{figure}

By using the value of $\Delta\hat{\alpha}$ and the noises of the CMB experiments, in Fig.\ref{figb3} we plot the residual BB power spectrum for the cases of $|\alpha|=4.32^{\circ}$ and $|\alpha|=0.1^{\circ}$, respectively. In the former case, we find that the residual spectra become more than two orders smaller than the original one, if the de-rotation is proceeded by considering the noises of the CMBPol or the reference experiment. In addition, they are all much smaller than the residual BB power spectrum caused by the cosmic weak lensing. In the latter case, the residuals are even smaller, which are entirely negligible, comparing with the residual BB spectrum of weak lensing.

Now, we repeat the calculation of signal-to-noise ratio for the gravitational-wave detection in Sec. \ref{sec2}. But here, we consider the residual BB power spectrum, instead of the total rotated BB spectrum, as the contamination. The results are presented in Fig.\ref{figb4}, where the dashed line denotes the result of the case where the Planck noises are considered for the de-rotating, the dash-dotted line is that for the CMBPol noise case, and the dotted line is for the case with noises of the reference experiment. We find that, for the Planck noise case, the decreasing of $r_{\rm min}$ is small. This is because that the Planck noises are relatively high, and the reconstruction of the B-mode is quite weak, which is consistent with the results in Fig.\ref{figb3}. However, if the noises of CMBPol or the reference experiment are considered for the de-rotating, the values of $r_{\rm min}$ are significantly reduced. Even if the rotation angle is $|\alpha|=4.32^{\circ}$, we have $r_{\rm min}=1.5\times10^{-6}$ for CMBPol case, and $r_{\rm min}=1.7\times10^{-7}$ for the reference experiment case. We should mention that, the residual BB power spectrum caused by cosmic weak lensing also follows a detection limit $r_{\rm min}=1.5\times10^{-5}$ \cite{knox2002,kamionkowski2002,zhao2014}. Comparing with them, we conclude that, if de-rotating the B-mode spectrum by considering the noise level of CMBPol mission or the better experiment, the residual BB power spectrum becomes completely negligible for the detection of relic gravitational waves in the CMB.

\section{Conclusions \label{sec4}}

The cosmological birefringence caused by the Chern-Simons coupling of the cosmic scalar field to the electromagnetic field has the possibility to rotate the polarization planes of the CMB photons when they propagate from the last scattering surface to us and to convert a part of  E-mode polarization to the B-mode polarization. Such kind of rotated BB power spectrum at late time forms a new contamination for the detection of relic gravitational waves in the CMB. If the rotation angle $\alpha$ is close to the current upper limit value, we find that the gravitational-wave detection is limited by this new noise if the tensor-to-scalar ratio is smaller than $0.0014$.

In this paper, we suggest the method to partly reconstruct and subtract the rotated B-mode polarization by utilizing the statistical properties of the estimators of E-mode coefficients $E_{\ell m}$ and the rotation angle $\alpha$. We find that, if this de-rotating is done by considering the noise level of the CMBPol mission or the better experiments, the residual BB power spectrum can be reduced by more than two orders, even if the largest rotation angle is considered. The residuals are much smaller than the de-lensed BB power spectrum caused by the cosmic weak lensing, and become negligible for the detection of gravitational waves.

~

~

~

{\it Acknowledgments:}
W.Z. is supported by project 973 under Grant No.2012CB821804, by NSFC No.11173021, 11322324 and project of KIP of CAS. M.L. is supported by
Program for New Century Excellent Talents in University and by NSFC under Grants No. 11075074.

\appendix


\begin{thebibliography}{99}

\bibitem{grishchuk1979}
L. P. Grishchuk, Sov. Phys. JETP {\bf 40}, 409 (1975), Ann NY Acad. Sci. {\bf 302}, 439 (1977), JETP Lett. {\bf 23}, 293 (1976).

\bibitem{starobinsky1980}
A. A. Starobinsky, JETP lett. {\bf 30}, 682 (1979).

\bibitem{turner1994}
L. Knox and M. S. Turner, Phys. Rev. Lett. {\bf 73}, 3347 (1994).

\bibitem{zhao2009}
W. Zhao, Phys. Rev. D {\bf 79}, 063003 (2009).

\bibitem{zaldarriaga1997}
U. Seljak and M. Zaldarriaga, Phys. Rev. Lett. {\bf 78}, 2054 (1997).


\bibitem{kamionkowski1997}
M. Kamionkowski, A. Kosowsky and A. Stebbins, Phys. Rev. Lett. {\bf 78}, 2058 (1997).


\bibitem{task}
J. Bock et al. astro-ph/0604101.

\bibitem{krauss2010}
L. M. Krauss, S. Dodelson, and S. Meyer, Science {\bf 328},
989 (2010).

\bibitem{wmap9}
G. F. Hinshaw {\it et.al.} [WMAP Collaboration], ApJS  {\bf 208}, 19 (2013).

\bibitem{planck}
P.~A.~R.~Ade {\it et al.}  [Planck Collaboration],
  arXiv:1303.5076.


\bibitem{bicep1}
H. C. Chiang, P. A. R. Ade, D. Barkats {\it et al.} Astrophys. J. {\bf 711}, 1123 (2010);
D. Barkats {\it et al.} [BICEP1 Collaboration], Astrophys. J. {\bf 783}, 67 (2014).

\bibitem{sptpol}
D. Hanson {\it et al.} [SPT Collaboration], Phys. Rev. Lett. {\bf 111}, 141301 (2013).

\bibitem{polarbear}
P. A. R. Ade {\it et al.} [POLARBEAR Collaboration], Phys. Rev. Lett. {\bf 112}, 131302 (2014).


\bibitem{actpol}
S. Naess {\it et al.} [ACT Collaboration], arXiv:1405.5524.

\bibitem{bicep2}
P. A. R. Ade {\it et al.} [BICEP2 Collaboration], Phys. Rev. Lett. {\bf 112}, 241101 (2014).



\bibitem{debate1}
M. J. Mortonson and U. Seljak, arXiv:1405.5857.

\bibitem{debate2}
R. Flauger, J. C. Hill, and D. N. Spergel, arXiv:1405.7351.

\bibitem{liuhao}
H. Liu, P. Mertsch and S. Sarkar, Astrophys. J. {\bf 789}, L29 (2014).

\bibitem{c1}
F. Schmidt and D. Jeong, Phys. Rev. D {\bf 86}, 083513 (2012).

\bibitem{c2}
F. Schmidt, E. Pajer and M. Zaldarriaga, Phys. Rev. D {\bf 89}, 083517 (2014).


\bibitem{d1}
S. Dodelson, Phys. Rev. D {\bf 82}, 023522 (2010).


\bibitem{c3}
N. E. Chisari, C. Dvorkin and F. Schmidt, arXiv:1406.4871.


\bibitem{zaldarriaga1998}
M. Zaldarriaga and U. Seljak, Phys. Rev. D {\bf 58}, 023003 (1998).

\bibitem{lewis2006}
A. Lewis and A. Challinor, Phys. Rep. {\bf 429}, 1 (2006).

\bibitem{carroll} S. M. Carroll, Phys. Rev. Lett. {\bf 81}, 3067
(1998).

\bibitem{kamionkowski1998}
A. Lue, L. Wang and M. Kamionkowski, Phys. Rev. Lett. {\bf 83} 1506 (1999).

\bibitem{other-cpt}
N.~F. Lepora, arXiv: gr-qc/9812077;
K.~R.~S. Balaji, R.~H. Brandenberger and D.~A. Easson, JCAP {\bf 0312}, 008 (2003);
B. Feng, H. Li, M. Li and X. Zhang, Phys. Lett. B {\bf 620}, 27 (2005);
M. Li, J. -Q. Xia, H. Li and X. Zhang, Phys. Lett. B {\bf 651}, 357 (2007);
C.~Q.~Geng, S.~H.~Ho and J.~N.~Ng,
  JCAP {\bf 0709}, 010 (2007);
Y.~-F.~Cai, M.~Li and X.~Zhang,  JCAP {\bf 1001}, 017 (2010).

\bibitem{li2009}
M. Li, Y. -F. Cai, X. Wang, and X. Zhang, Phys. Lett. B {\bf 680}, 118 (2009).

\bibitem{Faraday1}
  A.~Kosowsky and A.~Loeb,
  Astrophys.\ J.\  {\bf 469}, 1 (1996).

\bibitem{Faraday2}
  A.~Kosowsky, T.~Kahniashvili, G.~Lavrelashvili and B.~Ratra,
  Phys.\ Rev.\ D {\bf 71}, 043006 (2005).

\bibitem{Faraday3}
  M.~Giovannini,
  arXiv:1404.3974 [astro-ph.CO].

\bibitem{feng2006}
B. Feng, M. Li, J. Xia, X. Chen and X. Zhang, Phys. Rev. Lett. {\bf 96}, 221302 (2006).


\bibitem{other-constraints}
G. -C. Liu, S. Lee and K. -W. Ng, Phys. Rev. Lett. {\bf 97}, 161303 (2006);
A. Kostelecky and M. Mewes, Phys. Rev. Lett. {\bf 99}, 011601 (2007);
P. Cabella, P. Natoli and J. Silk, Phys. Rev. D {\bf 76}, 123014 (2007);
J. -Q. Xia, H. Li, X. Wang and X. Zhang, A\&A {\bf 483}, 715 (2008);
J. -Q. Xia, H. Li, G. -B. Zhao and X. Zhang, Astrophys. J. {\bf 679}, L61 (2008);
T.~Kahniashvili, R.~Durrer and Y.~Maravin,
  Phys.\ Rev.\ D {\bf 78}, 123009 (2008);
  V.~A.~Kostelecky and M.~Mewes,
  Astrophys.\ J.\  {\bf 689}, L1 (2008);
E.~Y.~S.~Wu {\it et al.}  [QUaD Collaboration],
  Phys.\ Rev.\ Lett.\  {\bf 102}, 161302 (2009);
E. Komatsu {\it et al}., Astrophys. J. Suppl. {\bf 180}, 330 (2009);
L. Pagano {\it et al}., Phys. Rev. D {\bf 80}, 043522 (2009);
J.~-Q.~Xia, H.~Li and X.~Zhang,
  Phys.\ Lett.\ B {\bf 687}, 129 (2010);
  S.~d.~S.~Alighieri, F.~Finelli and M.~Galaverni,
  Astrophys.\ J.\  {\bf 715}, 33 (2010);
E. Komatsu {\it et al}., Astrophys. J. Suppl. {\bf 192}, 18 (2011);
A.~Gruppuso, P.~Natoli, N.~Mandolesi, A.~De Rosa, F.~Finelli and F.~Paci,
  JCAP {\bf 1202}, 023 (2012);
  G.~Gubitosi and F.~Paci,
  JCAP {\bf 1302}, 020 (2013);
  {\tb{S.Y Li, J.Q. Xia, M. Li, H. Li and X. Zhang, arXiv:1405.5637.}}


\bibitem{xia2012}
J. -Q. Xia, JCAP {\bf 1201}, 046 (2012).

\bibitem{li2008}
M. Li and X. Zhang, Phys. Rev. D {\bf 78}, 103516 (2008).


\bibitem{cpt-perturbation}
M. Pospelov, A. Ritz and C. Skordis, Phys. Rev. Lett. {\bf 103}, 051302 (2009);
M. Kamionkowski, Phys. Rev. Lett. {\bf 102}, 111302 (2009);
A. P.S. Yadav, R. Biswas, M. Su and M. Zaldarriaga, Phys. Rev. D {\bf 79}, 123009 (2009);
R. R. Caldwell, V. Gluscevic and M. Kamionkowski, Phys. Rev. D {\bf 84}, 043504 (2011);
V. Gluscevic, D. Hanson, M. Kamionkowski and C. M. Hirata, Phys. Rev. D {\bf 86}, 103529 (2012);
M. Li and B. Yu, JCAP {\bf 1306}, 016 (2013);
S. Lee, G. -C. Liu and K. -W. Ng, Phys. Rev. D {\bf 89}, 063010 (2014);
{\tb{S. Lee, G. -C. Liu and K. -W. Ng, arXiv:1403.5585.}}

\bibitem{kamionkowski2009}
V. Gluscevic, M. Kamionkowski and A. Cooray, Phys. Rev. D {\bf 80}, 023510 (2009); A. Z. Wang, Q. Wu, W. Zhao and T. Zhu, Phys. Rev. D {\bf 89}, 103518 (2013).


\bibitem{zhao2014}
{\tb{W. Zhao and M. Li, Phys. Rev D {\bf 89}, 103518 (2014).}}

\bibitem{carroll1990}
S. M. Carroll, G. B. Field and R. Jackiw, Phys. Rev. D {\bf 41}, 1231 (1990).

\bibitem{Myers:2003fd}
  R.~C.~Myers and M.~Pospelov,
  Phys.\ Rev.\ Lett.\  {\bf 90}, 211601 (2003)

\bibitem{Kostelecky:2009zp}
  V.~A.~Kostelecky and M.~Mewes,
  Phys.\ Rev.\ D {\bf 80}, 015020 (2009).



\bibitem{zhao20092}
W. Zhao, D. Baskaran and L. P. Grishchuk, Phys. Rev. D {\bf 79}, 023002 (2009); Phys. Rev. D  {\bf 80}, 083005 (2009); Phys. Rev. D  {\bf 82}, 043003 (2010);
W. Zhao and D. Baskaran, Phys. Rev. D {\bf 79}, 083003 (2009).


\bibitem{lewis2002}
A. Lewis and S. Bridle, Phys. Rev. D {\bf 66}, 103511 (2002).



\bibitem{fisher}
M. Tegmark, A. Taylor and A. Heavens, Astrophys. J. {\bf 480}, 22 (1997);
M. Zaldarriaga, D. Spergel and U. Seljak, Astrophys. J. {\bf 488}, 1 (1997).


\bibitem{planck-bluebook}
J.~Tauber {\it et al.}  [Planck Collaboration],
  astro-ph/0604069.

\bibitem{ma2010}
Y. -Z. Ma, W. Zhao and M. Brown, JCAP {\bf 1010}, 007 (2010).

\bibitem{cmbpol}
CMBPol Study Team collaboration, D. Baumann et al., AIP Conf. Proc. {\bf 1141}, 10 (2009).


\bibitem{zhao2011}
W. Zhao, JCAP {\bf 1103}, 007 (2011).


\bibitem{hu2002}
W. Hu and T. Okamoto, Astrophys. J. {\bf 574} 566 (2002).


\bibitem{knox2002}
L. Knox and Y. -S. Song, Phys. Rev. Lett. {\bf 89}, 011303 (2002).


\bibitem{kamionkowski2002}
M. Kesden, A. Cooray and M. Kamionkowski, Phys. Rev. Lett. {\bf 89}, 011304 (2002).




\end{thebibliography}
\end{document}